\documentstyle[sprocl]{article}

\newcounter{saveeqn}%
\newcommand{\alpheqn}{\setcounter{saveeqn}{\value{equation}}%
\stepcounter{saveeqn}\setcounter{equation}{0}%
\renewcommand{\theequation}
   {\mbox{\arabic{saveeqn}\alph{equation}}}}%
\newcommand{\reseteqn}{\setcounter{equation}{\value{saveeqn}}%
\renewcommand{\theequation}{\arabic{equation}}}%

\def\ra{\rightarrow}

\def\be{\begin{equation}}
\def\ee{\end{equation}}
\def\bea{\begin{eqnarray}}
\def\eea{\end{eqnarray}}
\def\tb{\tilde{\beta}}

\begin{document}
\renewcommand{\baselinestretch}{2.0}
\vspace*{-1in}
\smallskip
\hspace{3in} CMU-HEP-98-01


\title{New Physics Effects in CP-Violating B Decays}

\author{L. WOLFENSTEIN}
\address{Department of Physics, Carnegie Mellon University, Pittsburgh,\\ Pennsylvania  15213, USA}

\setlength{\baselineskip}{36pt}

\maketitle

\bigskip

\abstracts{Contributions to $B - \bar B$ mixing from physics beyond the standard model may be
detected from CP-violating asymmetries in B decays.  There exists the possibility of large new
contributions that cannot be detected by first generation experiments because of a discrete
ambiguity.  Some possible strategies for resolving this are discussed.}

A major goal of the experiments on B mesons is to check the standard model, or conversely, to
discover new physics.  In many models beyond the standard model, there exist new contributions to
$ B - \bar B$ mixing [1]. 
 In this paper, we assume that
this is the only new physics and discuss strategies to detect it.  An
important conclusion is that even large new contributions due to $ B_d -
\bar{B_d}$ mixing may be difficult to detect.  Of course in some models,
the existence of such large new contributions might imply other
deviations from the standard model such as the rates for rare decay
processes [2].

The first information on $ B - \bar B$ mixing comes from the measurement
of $\Delta \ m$ or $x_d = 
\Delta \ m / \Gamma$.  This is proportional to 

$$ A^2 \left [ (1 - \rho)^2 + \eta^2) \right]  B_B \ \eta_2 \ f^2_B  $$

\noindent where $B_B \ \eta_2 \ f^2_B$ involves the hadronic matrix element.  Given the hadronic
uncertainty and conservative limits on the CKM matrix parameters ($\rho , \eta$) the standard model predicts
$x_d$ only within a factor of about ten.  The experimental result $x_d = 0.7$ fits very nicely but
provides weak constraints on new physics.

The next information, a major goal of B factories, is the phase of $M_{12}$ in the standard phase
convention.  This is given by $2 \tb$ and determined from measuring the CP-violating asymmetry
$\rm{sin} 2 \tb$ in the decay $B \ra \psi K_s$.  In the standard model $\tb 
= \beta $, the phase of
$V_{td}$, and is constrained to lie between 8$^{\circ}$ and 32$^{\circ}$
corresponding to $\rm{sin} \
 2 \beta$ between 0.3 and 0.9.  Thus a magnitude clearly below 0.3 or a negative value of $\rm{sin} \ 2
\tb$ would indicate new physics.

To proceed we assume that measurements yield $\rm{sin} 2 \tb$ between 0.4 and 0.8 corresponding in the
standard model to a value

$$ \tb \ = \tb_1 \ = \ 12 ^{\circ} \ \rm{to} \ 27 ^{\circ} $$

\noindent There exists the possibility that the true value of $\tb$ is 

\alpheqn
\bea
 \tb_2 \ = \frac{\pi}{2} - \tb_1 , \ = \ 78 ^{\circ} \ \rm{to} \ 63
^{\circ}  
\eea

This would mean a large new physics contribution that reverses the sign of $R
e \ M_{12}$.  Within the standard phase convention this new physics
contribution could be approximately CP invariant.  As we now proceed to
show this large new physics effect is not easy to detect.

The next goal of B factories is the measurement of $\rm{sin} \ 2 \ ( \ \tb + \gamma \ )$ from the
asymmetry in decays like $B^0 \ra \pi^+ \pi^-$.  For the moment we neglect the penguin problem and
assume this is measured.  In the standard model there is almost no
constraint [3]
 on the possible value of $ \rm{sin} \ 2 \ ( \ \tb + \gamma \ )$ for a value
of $\rm{sin} \ 2
\tb$ in the range we have assumed.  Within the standard model there will in general be only one
set of angles $( \ \tb_1 , \gamma_1 \ )$ consistent with these two measurements, although in
general there is an eight-fold ambiguity [4].
In particular, corresponding to the choice $\tb  = \tb_2$ there is a
corresponding choice

\bea
\gamma_2 \ = \ \pi - \gamma_1 
\eea
\reseteqn

Since the allowable values of $\gamma$, which are independent of $B - \bar B$ mixing and in
our scenario are unchanged by the new physics, are approximately symmetric with respect to 90 
 $^{\circ}$  the choice $\gamma_2$ is always allowable.  A number of experiments are directed at
determining $\rm{sin} \gamma$; this does not distinguish $\gamma_1$, from $\gamma_2$.

If $\gamma_1$ is far from 90  $^{\circ}$ corresponding to $\left| \rho \right| \ \geq \ 0.2$ then
$\gamma_2$ is distinguished from $\gamma_1$ by the sign of $\rho$ and thus by the magnitude of
$V_{td}$.  The best prospect for determining this is from the rate
[5]
of $K^+ \ra \pi^+ \ \nu \bar \nu $ which is approximately proportional
to 

$$ \left[ \ (1 \pm .15) \ + \ (2 \ \pm .25) \ (1 - \rho - i \ \eta) \
\right]^2 $$

\noindent where the first conservative error is due to the charm
contribution and the second to uncertainty in
$m_t$ and $V_{cb}$.  For $\left| \rho \right| \ = \ 0.2$ the difference
between the two signs of $\rho $ is almost a factor of 2 in the $K^+ \ra
\pi^+ \nu \bar \nu$ rate.

Another possibility is to look for interfering amplitudes that can be used to determine $\rm{cos}
\gamma$.  An example is the penguin-tree interference in the decay $B^0 \ra \pi^- K^+ $.  In
contrast one expects that the decay $B^+ \ra \pi^+ K^0 $ is pure penguin.  One then finds [6]

\be 
  R \ = \ \frac{\Gamma \ ( \ B^0 \ra \pi^- K^+ \ )}{\Gamma \ ( \ B^+ \ra \pi^+ K^0 \ )} \ = \ 1 \
- \ 2 r \ \rm{cos} \ \gamma \ + r^2
\ee

\noindent where $r$ is the ratio of tree to penguin.  If we accept the sign of $r$ as given by
factorization and note that we expect $r \ \leq \ \frac{1}{3}$ then the sign of $(1-R)$ gives the
sign of $\rm{cos} \  \gamma$ which can distinguish $\gamma_1$ from $\gamma_2$.

However, if $\rm{cos} \  \gamma$ is close to zero, corresponding to $\rho$ close to zero, which is
in the center of the allowed $(\rho , \ \eta)$ region, then neither of the above methods
can distinguish the solutions in Eq. (2) from the standard model.

Instead of relying on $\gamma $ one can try to find a method of distinguishing $\tb_1$ from
$\tb_2$.  Grossman and Quinn [4]
suggest comparing the asymmetry in the decay $B \ra D^+ D^-$ to that of $B \ra \psi \ K_s$.  Including
a penguin contribution to the $D^+ D^-$ decay they find

\be 
 a \ ( D^+ D^- ) \ = \ \rm{sin} \ 2 \tb \ - \ 2 r \ \rm{cos} \ 2 \tb
\ \rm{sin} \ \tb \ \rm{cos} \ \delta
\ee

\noindent where $r$ is the penguin to tree amplitude ratio and $\delta $ is the strong phase difference
between penguin and tree.  If one assumes $r < 0$ from factorization and $\rm{cos} \ \delta > 0 $ then
if $\tb \ = \  \tb_1 $ the asymmetry is reduced due to the penguin whereas if $\tb \ = \  \tb_2$
the asymmetry is increased.

Actually if $\tb \ = \  \tb_2$ Eq. (3) is not correct since it assumes that the phase of the
penguin amplitude, given by the phase $\beta $ of $V_{td}$, equals $\tb $. 
However in the scenario we consider while $\tb $ is given by Eq. (1a), the phase $\beta $ is
constrained to lie between 12$^{\circ} $ and 27$^{\circ} $.  In this case Eq. (3) becomes
to first order in $r$ 

\be 
a \ ( D^+ D^- ) \ = \ \rm{sin} \ 2 \tb_2 \ - \ 2 r \ \rm{cos} \ 2 \tb_2
\ \rm{sin} \ (2 \ \tb_2 - \beta) \ \rm{cos} \ \delta
\ee

\noindent The previous conclusion that if $ r < 0$ the asymmetry is increased by the penguin if
$\tb \ = \ \tb_2 $ still holds.

Another way to directly distinguish $\tb_2 $ from $\tb_1 $ in this scenario involves
decays dominated by the $b \ra d$ penguin graph.  Assuming $t$ dominance the asymmetry of a decay
like $B_d \ra K^0 \overline{K^0} $ is given by $\rm{sin} \ 2 \ (\tb - \beta) $.  If we assume $\tb $
is around 70$^{\circ}$, corresponding to typical $\tb_2 $ value then any allowable value of
$\beta$ gives an asymmetry greater than 0.9.  In contrast in the standard model $\tb \ = \ \beta $
and the asymmetry vanishes.  Fleischer [7] 
has pointed out that there may be significant contributions from u and c quarks such that the standard
model value may not be zero.  Nevertheless a very large asymmetry of 80\% or greater would be strong
evidence for new physics.  While the branching ratio is small not so many events are needed just to
show that the asymmetry is very large.

We turn now to the $B_s$ system.  The first quantity of interest that can be measured is
$x_s$.  The ratio $x_d$/$x_s$ is given in the SM by

\be 
\frac{x_d}{x_s} \ = \ \lambda^2 \ \left[ \ ( \ 1 - \rho \ )^2 \ + \ \eta^2 \ \right] \ K
\ee

\noindent where $K$ is the ratio of $B_B \ \eta_2 \ f^2_B $ for the $B_d$ as compared to $B_s$.  In
the SU(3) limit $K = 1$ and estimates from lattice and other calculations give $K$ between 0.7 and
0.9.  Thus the measurement of $x_s $ can be used to put a constraint on $( \rho , \ \eta ) $,
primarily on
$\rho$.  In fact the present limit on $x_s $ disfavors values $\rho < -0.2 $.  A small value
of $x_s $ leads to a significant negative value of $\rho $ and a large value of
$x_s$ to a positive value $\rho $.  If this is inconsistent with the value of $(\rho , \ \eta)$
determined from the asymmetry measurements it could be a sign of new physics in $B_d -
\overline{B_d}
$ mixing.  Note that this new physics in general would cause $\tb$ to be different from $\beta$ and
change the value of $x_d $ invalidating Eq. (5).  However, the larger new contribution to
$B_d - \overline{B_d} $ mixing implied by Eq. (1a) could not be demonstrated in this way.

It would also be possible to compare the values of $( \rho , \ \eta ) $, mainly $(1 - \rho) $,
that fits
$x_s / x_d$ with that from $ K^+ \ra \pi^+   \nu \bar{\nu} $.  If these are inconsistent
it would be probably a sign of a new physics contribution to $x_d$.

If $\Delta  m_s $ is not too large one can study the CP violating asymmetries from the $\rm{sin}
\ ( \Delta  m_s \ t )$ term in tagged  $B_s$ decays.  For decays such as $B_s \ra \psi \ \eta $
the asymmetry is given by $\rm{sin} \ \theta_s $ where $\theta_s  =  2 \ \lambda^2 \ \eta$
which is between .02 and .05.  If the asymmetry is significantly larger that would be a sign of
new physics in $B_s - \overline{B_s} $ mixing.  For decays governed by $b \ra u  \bar u d$, such
as $ B_s \ra
\rho^0  \  K_S $, the asymmetry in the tree approximation is $\rm{sin} \  ( \theta_s  +  2 \
\gamma ) $.  If $\theta_s$ is consistent with zero this gives $\rm{sin} \ 2 \ \gamma$, the sign of
which distinguishes $\gamma_2$ from $\gamma_1$.  There is likely a sizable penguin contribution to $B_s
\ra \rho^0 \ K_s$, but the fact that one wants only the sign of $\rm{sin} \ \gamma $ may make this
useful in spite of the penguin.

In analyzing prospective B asymmetry experiments its is natural and appropriate to assume the
standard model and see how well these can constrain the parameters  $( \rho , \ \eta ) $.  The
purpose of the present note is to emphasize that it is also important to look at new physics
effects and see whether or not a given set of experiments can detect them.

In particular we have looked at one particular ambiguity given by Eqs. (1), which implies large
new physics effects which may prove very difficult to
\newpage

\noindent detect.  Proposed experiments should be analyzed from the
point of view of resolving such ambiguities.

\section*{Acknowledgments} This research was supported in part by the U.S. Dept. of Energy under
DE-FG02-91-ER-40682.

\end{document}